\documentstyle[prl,aps,twocolumn,psfig]{revtex}
\newcommand{\ber}{\begin{eqnarray}}
\newcommand{\eer}{\end{eqnarray}}
\begin{document}
\twocolumn[\hsize\textwidth\columnwidth\hsize\csname
@twocolumnfalse\endcsname
\hfill LBL-46896
\title{
Kinetic equation with exact charge conservation
}
\author{
C.M. Ko$^a$, V. Koch$^b$, Zi-wei Lin$^a$, K. Redlich$^{b,c}$, 
M. Stephanov$^{d,e}$, and Xin-Nian Wang$^b$
}
\address{
        $^a$Cyclotron Institute and Physics Department,
Texas A\&M University, College Station, Texas 77843-3366
}
\address{
        $^b$Nuclear Science Division,
Lawrence Berkeley National Laboratory, 1 Cyclotron Road,
Berkeley, CA 94720}
\address{
        $^c$Institute of Theoretical Physics, University of
Wroc\l aw, PL-50204 Wroc\l aw, Poland}
\address{
         $^d$Department of Physics, University of Illinois,
         Chicago, IL 60607-7059
}
\address{
         $^e$RIKEN-BNL Research Center, Brookhaven
         National Laboratory, Upton, NY 11973
}
\date{\today}
\maketitle

\begin{abstract}
We formulate the kinetic master equation describing the production of
charged particles which are created or destroyed only in pairs
due to the conservation of their Abelian charge.
Our equation applies to arbitrary particle multiplicities and
reproduces the equilibrium results
for both canonical (rare particles) and grand canonical
(abundant particles) systems.
For canonical systems, the equilibrium multiplicity is much
lower and the relaxation time is much shorter than the
naive extrapolation from the grand canonical ensemble results.
Implications for particle chemical equilibration in heavy-ion collisions are
discussed.
\vspace{0.5in}
\end{abstract}


]

\section{Introduction}

Relativistic statistical thermodynamics has long been used 
as a tool to describe particle production in heavy-ion and
in high-energy particle collisions \cite{Hag71,sur,he}.
Recent analyses have shown that the statistical models
can indeed give a satisfactory description of the multiplicities
of most hadrons measured in A-A collisions at AGS and SPS
energies\cite{Bra99,our}.
However, the dynamics of particle equilibration, and in particular
chemical equilibration, is still not well understood
\cite{brat,bra,brown}.
In this Letter, we shall address the problem of chemical
equilibration within the statistical kinetic approach.

Within this approach, particle production 
is commonly described using the grand canonical ensemble,
where event-averaged multiplicities are controlled
by chemical potentials. In this description the net value of
a given $U(1)$ charge
(e.g., electric, baryon, strangeness, charm, etc.)
fluctuates. These fluctuations can be neglected
only if the particles carrying the charge in question are
abundant. In this case the charge will be conserved on the average.
In the opposite limit of rare particle production, conservation laws
must be implemented locally on an event-by-event basis \cite{Hag71,sur,all,Hag85};
i.e.,a canonical ensemble must be used.

The local conservation of quantum numbers in the canonical
approach severely reduces the phase space available
for particle production \cite{Hag71,sur,our,all,Hag85,Raf80,Cle91}.
Recently, it has been shown that the canonical statistical model
provides a good description of particle yields measured
in low-energy heavy-ion \cite{Cle99} and high-energy hadron-nucleus,
hadron-hadron
and $e^+e^-$ reactions \cite{our,mul,Bec96}.

In view of these results it is of particular importance to formulate
a kinetic theory for the time evolution of particle production
in order to investigate the approach to the canonical chemical
equilibrium. Obviously, if particles are abundantly produced, the 
equilibrium result for the particle multiplicity
should coincide with the grand canonical one.
For rare processes, however, the particle production
is strongly correlated and the canonical equilibrium result is expected.
The kinetic formulation for the production of strongly correlated
particles was studied in the literature \cite{koch,tur}.
However, no complete solution has been obtained.

In this letter we consider the time evolution of the multiplicity
of particles that carry the charges corresponding to an $U(1)$
internal symmetry.
We formulate a kinetic master equation valid for arbitrary particle
multiplicity.
It reproduces the canonical equilibrium solution for rare particle
production,
and reduces to the standard grand canonical rate equation for abundant
particle production.

\section{Rate equation}
In the standard formulation \cite{koch,rm,mat,rate},
the rate equation for a binary process $a_1 a_2 \rightarrow b_1 b_2$
with $a \neq b$ is described by the following population equation:
\begin{eqnarray}
\frac{d\langle N_{b_1}\rangle}{d\tau}={G\over V} 
\langle N_{a_1}\rangle \langle N_{a_2}\rangle 
- {L\over V} \langle N_{b_1}\rangle \langle N_{b_2}\rangle,
\label{normal1}
\end{eqnarray}
where $G \equiv \langle \sigma_G v \rangle$ and
$L \equiv \langle \sigma_L v \rangle$
give the momentum-averaged cross sections
for the gain process $a_1 a_2 \rightarrow b_1 b_2$
and the loss process $b_1 b_2 \rightarrow a_1 a_2$, respectively.
$N_k$ represents the total number of particles $k$,
and $V$ is the proper volume. Among such processes, a typical example 
is the kaon production/annihilation via $\pi^+\pi^-\leftrightarrow
K^+K^-$. The above rate equation, however, cannot be applied to the
situation where particle production is rare and is strongly correlated
by exact charge conservation. 

To account for the correlation between the production/annihilation 
of particles $b_1$ and $b_2$, let us define $P_{i,j}$ as the probability 
to find a number $i$ of particle $b_1$ and a number $j$ of particle 
$b_2$ in an event. We also denote as $P_i$ the probability to find a 
number $i$ of particle $b$ in an event.
The  average number of particles $b$ per event is then defined as:
\begin{eqnarray}
\langle N_b \rangle =\sum_{i=0}^{\infty} iP_i.
\label{2}
\end{eqnarray}

We can now write the following general rate equation for the average
particle multiplicity:
\begin{eqnarray}
\frac{d\langle N_{b_1} \rangle }{d\tau}=
{G\over V} \langle N_{a_1} \rangle
\langle N_{a_2} \rangle - \frac {L}{V}
\sum_{i,j} ij P_{i,j}.
\label{general}
\end{eqnarray}

Further, let particles $b_1$ and $b_2$ carry opposite units of a charge,
corresponding to an $U(1)$ internal symmetry (strangeness in the case of kaons).
Then the $U(1)$ charge neutrality of the system
gives $N\equiv N_{b_1}=N_{b_2}$. We have then,
\begin{eqnarray}
P_{i,j}&=& P_i ~\delta_{ij}, \nonumber\\
\sum_{i,j} ij P_{i,j} &=& \sum_i i^2 P_i
\equiv \langle N^2 \rangle =\langle N \rangle ^2+\langle \delta N^2
\rangle,
\label{5}
\end{eqnarray}
where $\langle \delta N^2 \rangle$ represents
the event-by-event fluctuation of the number of $b_1 b_2$ pairs.
Note that we always consider $a_1$ and $a_2$ particles abundant (such
as, e.g., pions) so that
we can neglect the event-by-event fluctuations of their multiplicity
and the change of their number due to the considered processes.

Following Eqs.(\ref{general}) and (\ref{5}), the general rate equation
for the average number of $b_1b_2$ pairs can be written as
\begin{eqnarray}
\frac{d\langle N \rangle }{d\tau}={G\over V}
\langle N_{a_1} \rangle \langle N_{a_2} \rangle - \frac
{L}{V}
\langle N^2 \rangle .
\label{general2}
\end{eqnarray}
For abundant production of $b_1 b_2$ pairs, 
where $\langle N \rangle \gg 1$,
\begin{eqnarray}
\langle N^2 \rangle \approx \langle N\rangle ^2,
\end{eqnarray}
and Eq.(\ref{general2}) obviously reduces to the standard form, i.e.,
\begin{eqnarray}
\frac{d\langle N\rangle }{d\tau}\approx \frac{G}{V}
\langle N_{a_1} \rangle \langle N_{a_2} \rangle
- \frac{L}{V} \langle N \rangle ^2.
\label{normal2}
\end{eqnarray}
However, for rare production of $b_1 b_2$ pairs,
where $\langle N\rangle\ll \!1$,
the rate equations (\ref{normal1}) and (\ref{normal2})
are no longer valid. We have instead
\begin{eqnarray}
\langle N^2 \rangle \approx \langle N\rangle ,
\end{eqnarray}
which reduces Eq.(\ref{general2}) to the following form:
\begin{eqnarray}
\frac{d\langle N\rangle }{d\tau}\approx {G\over V}
 \langle N_{a_1} \rangle \langle N_{a_2} \rangle
- \frac{L}{V} \langle N \rangle .
\label{canonical}
\end{eqnarray}
Thus, in the limit $\langle N \rangle \ll 1$,
the absorption term depends on the pair number
only {\it linearly}, instead of quadratically
for the limit $\langle N \rangle \gg 1$.

\section{Equilibrium multiplicity and relaxation time}
\label{limits}

To illustrate the differences in the time
evolution of particle abundance, we consider the two limiting cases,
$\langle N \rangle \gg 1$ and  $\langle N \rangle \ll 1$,
and present their equilibrium values and relaxation times for
the production of $b_1 b_2$ pairs.
As an example, we consider a system at fixed temperature and volume
and with no initial $b_1 b_2$ pairs, i.e.,
$\langle N \rangle (\tau=0) =0$.

In the limit when $\langle N \rangle \gg 1$,
the standard Eq.(\ref{normal2}) is valid and has the following well-known
solution:
\ber
\langle N \rangle ^{\rm GC} (\tau)=
N_{\rm eq}^{\rm GC} \tanh \left ( \tau/\tau_0^{\rm GC} \right ),
\eer
where the equilibrium value for the number of $b_1b_2$ pairs
$N_{\rm eq}^{\rm GC}$
and the relaxation time constant $\tau_0^{\rm GC}$ are given by
\ber
N_{\rm eq}^{\rm GC}= \sqrt \epsilon,~~~~
\tau_0^{\rm GC} = \frac {V}{L \sqrt \epsilon},
\label{eqgc}
\eer
respectively,
with $\epsilon\equiv G \langle N_{a_1} \rangle \langle N_{a_2}\rangle /L$.

In the particular case where particle momentum distributions are
thermal,
the gain ($G$) and loss ($L$) terms just represent the thermal averages
of the production and absorption cross sections with
\ber
\frac{G}{L}=
\frac {d_{b_1} \alpha_{b_1}^2 K_2 (\alpha_{b_1})
d_{b_2} \alpha_{b_2}^2 K_2 (\alpha_{b_2})}
{d_{a_1} \alpha_{a_1}^2 K_2 (\alpha_{a_1})
d_{a_2} \alpha_{a_2}^2 K_2 (\alpha_{a_2})},
\eer
where $d_k$'s denote the degeneracy factors, and $\alpha_k \equiv
m_k/T$.
The equilibrium value for the number of $b_1 b_2$ pairs
in Eq.(\ref{eqgc}) now reads as
\ber
N_{\rm eq}^{\rm GC}=
{ {d_{b_1}}\over {2\pi^2}} VT^3 \alpha_{b_1}^2 K_2 (\alpha_{b_1}).
 \label{15}
\eer
Thus it is described by the Grand Canonical (GC) result
with vanishing chemical potential due to our requirement of
the (average) $U(1)$ charge neutrality of the system.

In the opposite limit where $\langle N \rangle \ll 1$,
the time evolution is described by Eq.(\ref{canonical}),
which has the following solution:
\ber
\langle N \rangle ^{\rm C} (\tau)=
N_{\rm eq}^{\rm C} \left ( 1- e^{-\tau/\tau_0^{\rm C}} \right ),
 \label{16}
\eer
with  the equilibrium value and relaxation time  given by
\ber
N_{\rm eq}^{\rm C}= \epsilon,~~~~ \tau_0^{\rm C} = \frac {V}{L}.
\label{eqc}
\eer

With a thermal momentum distribution the equilibrium
value of $b_1b_2$ pair multiplicity has the following form:
\ber
N_{\rm eq}^{\rm C}= \left [ { {d_{b_1}}\over {2\pi^2}} VT^3
\alpha_{b_1}^2 K_2 (\alpha_{b_1}) \right ] \cdot 
\left [ { {d_{b_2}}\over {2\pi^2}} VT^3 \alpha_{b_2}^2 K_2
(\alpha_{b_2})
 \right ].
\label{lim}
\eer
This equation demonstrates the locality of the $U(1)$ charge
conservation.
With each particle $b_1$, a particle $b_2$ with the opposite charge
is produced in the same event in order to conserve charge
locally.
This is the result expected from the Canonical (C) formulation of
conservation laws \cite{Hag85,Raf80}.

We note that Eq.(\ref{lim}) is just the leading term in the expansion of
the canonical result for the multiplicity of particles carrying the $U(1)$
charges. The general expression is known to have the following form
\cite{Hag85,Raf80}:
\ber
N_{\rm eq}^{\rm C}= N_{\rm eq}^{\rm GC} {{I_1(2 N_{\rm eq}^{\rm
GC})}\over
{I_0(2 N_{\rm eq}^{\rm GC})}},
\label{19}
\eer
where
$ N_{\rm eq}^{\rm GC}$ is given by Eq.(\ref{15}) and $I_i$'s
are the modified Bessel functions.

Comparing Eq.(\ref{eqgc}) and Eq.(\ref{eqc}), we first find that,
for $\langle N \rangle \ll 1$, the equilibrium multiplicity
in the canonical formulation is much lower than
what is expected from the grand canonical result,
\ber
N_{\rm eq}^{\rm C}=(N_{\rm eq}^{\rm GC})^2 \ll {N_{\rm eq}^{\rm GC}}.
\eer
This shows the importance of the canonical description of charge
 conservation when the multiplicity of charged particles
is small.
We also note that the volume dependence in the two cases is different. 
The particle density in the GC (abundant) limit is independent of $V$,
whereas in the opposite canonical (rare) limit the density scales linearly with
$V$.

Secondly, the relaxation time for a canonical system is
far shorter than what is expected from the grand canonical result,
\ber
\tau_0^{\rm C}=\tau_0^{\rm GC} N_{\rm eq}^{\rm GC} \ll \tau_0^{\rm GC},
\eer
due to small number of particles ($N_{\rm eq}^{\rm GC}\ll 1$).
For example, the total number of produced kaons in $Au+Au$ collisions
at 1 GeV/$A$ is of the order of 0.02. Thus the canonical relaxation time
is a factor of 50 shorter than what is expected from the grand canonical
formulation.

We note from Eq.(\ref{general2})
that these two limits are essentially determined by the size of
$\langle \delta N^2 \rangle$, the event-by-event fluctuation
of the number of $b_1 b_2$ pairs.
The grand canonical results correspond to small fluctuations, i.e.,
$\langle \delta N^2 \rangle/\langle N \rangle^2 << 1$, while
the canonical description is necessary
in the opposite limit.

\section{Master equation}

In this section, we formulate the general evolution equation
which is valid for an arbitrary value of $\langle N \rangle$.
It is a master equation for $P_n(\tau)$,
the probability of finding $n$ pairs of $b_1b_2$ at time $\tau$.
This probability increases with time due to
transitions from $n-1$ and $n+1$ states to the $n$ state,
while it also decreases due to transitions from the $n$ state
to $n-1$ and $n+1$ states. The transition probability $n\to n+1$
per unit time due to pair creation is
 $G\langle N_{a_1} \rangle \langle N_{a_2}\rangle /V$ 
and the transition probability $n\to n-1$ due to pair annihilation
is $n^2 L/V$. Therefore, the master equation set
has the form:
\ber
\frac{dP_n}{d\tau}&=&{G\over V} \langle N_{a_1}
 \rangle \langle N_{a_2}\rangle\, \left [P_{n-1}-P_n \right] \nonumber
\\
&-& \frac {L}{V} \, \left [ n^2 P_n - (n+1)^2 P_{n+1} \right ],
\label{generaln}
\eer
where $n=0,1,2,3,\cdots$.
Multiplying the above equation by $n$ and summing over $n$,
one recovers Eq.(\ref{general2}), the general rate equation for the
time evolution of the average number of $b_1 b_2$ pairs. However,
the master equation (\ref{generaln}) contains much more information
than the rate equation (\ref{general2}). It contains enough
information to solve for the evolution of $\langle N\rangle$
(and all moments of $N$) for arbitrary $\langle N\rangle$.
For example, one can also obtain the time evolution of particle
fluctuations, $\langle \delta N^2\rangle$, which are of physical importance
for rarely produced particles.

We can convert the iterative equations (\ref{generaln}) 
for $P_n$'s into a partial
differential equation for the generating function
\begin{equation}
g(x,\tau) = \sum_{n=0}^\infty x^n P_n (\tau).
\label{23}
\end{equation}
Multiplying Eq.(\ref{generaln}) by $x^n$ and summing over $n$, we find
\ber
\frac {\partial g(x,\tau)}{\partial \tau} =
\frac {L}{V} (1-x) \left (x g''+ g'- \epsilon g \right ),
\label{gtau}
\eer
where $g' \equiv \partial g / \partial x$.
Note that the $g(1,\tau)$ does not change with time, which is
equivalent to the conservation of total probability evident in
Eq.(\ref{generaln}).

The equilibrium solution, $g_{\rm eq}(x)$, thus, obeys the following
equation:
\ber
xg_{\rm eq}'' + g_{\rm eq}' - \epsilon g_{\rm eq} = 0. 
\label{equil}
\eer
By variable substitution ($x = y^2/(4\epsilon)$) 
this equation can be reduced to
the Bessel equation. 
The solution that is regular at $x=0$ (since $g(0)=P_0 \leq 1$) is given by
\ber
g_{\rm eq}(x) = \frac {1}{I_0 ( 2\sqrt \epsilon)} I_0 ( 2\sqrt {\epsilon
x} ),
\label{26}
\eer
where the normalization is fixed by $g(1) = \sum P_n = 1$.

The equilibrium probability distribution $P_n$ can now be
found from Eqs.(\ref{23},\ref{26}) as
\ber
P_{n,\rm eq}=\frac {\epsilon^n} {I_0 ( 2\sqrt \epsilon )~(n!)^2},
\eer
and the value of the average number of $b_1 b_2$ pairs per
event at equilibrium is given by
\ber
\langle N \rangle_{\rm eq}= g'(1)=
\sqrt \epsilon ~\frac {I_1 ( 2\sqrt \epsilon)}{I_0 ( 2\sqrt \epsilon)},
\label{neq}
\eer
which obviously coincides with the expected result
for particle multiplicity in the canonical ensemble given by
Eq.(\ref{19})
and also reduces to the results for the two limiting cases in
Sec.\ref{limits}.

\section{Conclusions}

We have formulated the kinetic master equation for strongly correlated
production of particles,
where the correlation is due to
the local charge conservation required by an $U(1)$ internal symmetry.
Our general rate equation is valid for arbitrary value of $\langle N
\rangle$,
thus it reduces to the grand canonical results for large $\langle N
\rangle$
and to the canonical results for small $\langle N \rangle$.
Therefore, our equation provides a generalization of the standard rate
equation beyond the grand canonical limit.
We have shown that for rare particle production
the equilibrium multiplicity is much smaller and the relaxation time is
much shorter than expected from the standard rate equation.
For abundant particle production, where
the standard rate equation applies to first order,
one can use the general rate equations to
study finite-size corrections to the grand canonical results.

Our results could be of importance in the description and understanding
of equilibration phenomena and equilibrium properties of partonic or
hadronic
matter created in heavy-ion collisions.
For example, it could provide insights into the equilibration time of
strange particles produced at SIS, or open charm and charmonium
productions
at SPS and higher energies.
It may also be meaningful for transport model studies of rare particle
production, especially the perturbative procedure for rare processes
\cite{pert}, 
where the local charge conservation requires that 
the production probability should be assigned to the pair instead to 
each particle separately.

\section*{Acknowledgments}
\hspace*{\parindent}

We acknowledge stimulating discussions with M. Gyulassy,
S. Jeon and L.D. McLerran.
K.R. also acknowledges comments from C. Greiner, J. Knoll,
and H. Feldmeier.
M.S. and Z.L. are grateful to the nuclear theory group at LBNL
for supporting the summer program where this work was started.
This work was supported by the Director, Office of Energy Research,
Office of High Energy and Nuclear Physics, Division of Nuclear Physics,
the Office of Basic Energy
Science, Division of Nuclear Science, of the U.S. Department of Energy under
Contract No. DE-AC03-76SF00098,
the Committee for Scientific Research (KBN-2P03B 03018),
the National Science Foundation under Grant No. PHY-9870038,
the Welch Foundation under Grant No. A-1358,
and the Texas Advanced Research Program under Grant No.
FY99-010366-0081.

{}

\begin{thebibliography}{10}

\bibitem{Hag71} R. Hagedorn, CERN yellow report 71-12, 101 (1971).

\bibitem{sur} E.V. Shuryak,
Phys. Lett. {\bf B42}, 357 (1972); Sov. J. Nucl. Phys. {\bf 20}, 295 (1975).

\bibitem{he} U. Heinz,  Nucl. Phys. {\bf A661}, 349 (1999);
R. Stock, Phys. Lett. {\bf B456}, 277 (1999).

\bibitem{Bra99} P. Braun-Munzinger, I. Heppe and J. Stachel,
Phys. Lett. {\bf B465}, 15 (1999);
P.  Braun-Munzinger,  J. Stachel, J. P. Wessels and N. Xu,
Phys. Lett. {\bf B344}, 43 (1995); Phys. Lett. {\bf B365}, 1 (1996);
J.  Cleymans and K. Redlich,
Phys. Rev. Lett. {\bf 81}, 5284 (1998);
 D. Yen and M.I. Gorenstein,
Phys. Rev. {\bf C59} 2788 (1999);
 J. Letessier and J. Rafelski,
Int. J. of Mod. Phys.  {\bf E9} 107 (2000).

\bibitem{our}J. S. Hamieh, K. Redlich and A. Tounsi,
Phys. Lett. {\bf B486}, 61 (2000).

\bibitem{brat}J. E.L. Bratkovskaya, W. Cassing, C. Greiner,
M. Effenberger, U. Mosel and A. Sibirtsev,
Nucl. Phys. {\bf A675}, 661 (2000).

\bibitem{bra} L.V. Bravina, {\it et al.},
Phys. Lett. {\bf B459}, 660 (1999).

\bibitem{brown} G. E. Brown, M. Rho and C. Song (to be published).

\bibitem{all} K. Redlich and L. Turko,
Z. Phys.  {\bf B97}, 279 (1980), L. Turko,
Phys. Lett. {\bf B104}, 153 (1981), H.-Th. Elze,
W. Greiner and J. Rafelski, Phys. Lett. {\bf B124}, 515 (1983).

\bibitem{Hag85} R. Hagedorn and K. Redlich, Z. Phys. {\bf A27}, 541
(1985).

\bibitem{Raf80} J. Rafelski and M. Danos,
Phys. Lett. {\bf B97}, 279 (1980).

\bibitem{Cle91}J. Cleymans, K. Redlich and E. Suhonen,
Z. Phys. {\bf C51}, 137 (1991).

\bibitem{Cle99} J.  Cleymans and K. Redlich,
Phys. Rev. {\bf C60}, 054908 (1999);
J. Cleymans, H. Oeschler and K. Redlich,
Phys. Rev. {\bf C59}, 1663 (1999); Phys. Lett.
{\bf B485}, 27  (2000).


\bibitem{mul} B. M\"uller and J. Rafelski,
Phys. Lett. {\bf B116}, 274 (1982).
\bibitem{Bec96}
F. Becattini,  Z. Phys. {\bf C69}, 485 (1996);
F. Becattini and U. Heinz, Z. Phys. {\bf C76}, 269 (1997).

\bibitem{koch} P. Koch, B. M\"uller and J. Rafelski,
Phys. Rept. {\bf 142}, 167 (1986).

\bibitem{tur}  J. Rafelski and L. Turko,
hep-th/003079.

\bibitem{rm}  J. Rafelski and B. M\"uller,
Phys. Rev. Lett. {\bf 48}, 1066 (1982).

\bibitem{mat} T. Matsui, B. Svetitsky and L.D. McLerran,
Phys. Rev. {\bf D34}, 783 (1986).

\bibitem{rate} T.S. Biro, E. van Doorn, B. M\"uller,
M.H. Thoma and X.-N. Wang,
Phys. Rev. {\bf C48}, 1275 (1993).

\bibitem{pert}
X. S. Fang, C. M. Ko, G. Q. Li and Y. M. Zheng, 
Nucl. Phys. {\bf A575}, 766 (1994); C. M. Ko and G. Q. Li, J. Phys. 
{\bf G22}, 1673 (1996); W. S. Chung, G. Q. Li and C. M. Ko, Nucl. Phys. 
{\bf A625}, 347 (1997). 

\end{thebibliography}
\end{document}